\documentclass[aps,prd,superscriptaddress,showpacs]{revtex4}
\usepackage{graphicx}
\usepackage{epstopdf}
\usepackage{amsmath}
\usepackage{amsfonts}
\usepackage{amssymb}
\usepackage{latexsym}

\begin{document}
\title{Aspects of CPT-even Lorentz-symmetry violating physics in a supersymmetric scenario}
\author{H. Belich}
\email{belichjr@gmail.com}
\affiliation{Departamento de F\'{\i}sica e Qu\'{\i}mica, Universidade Federal do Esp\'{\i}rito Santo, Vit\'{o}ria, ES, Brasil}
\author{L. D. Bernald}
\email{ldurand@cbpf.br}
\affiliation{Centro Brasileiro de Pesquisas F\'\i sicas, Rio de Janeiro, RJ, Brasil}
\author{Patricio Gaete}
\email{patricio.gaete@usm.cl}
\affiliation{Departmento de F\'{\i}sica and Centro Cient\'{\i}fico-Tecnol\'ogico de Valpara\'{\i}so, Universidad T\'{e}cnica Federico Santa Mar\'{\i}a, Valpara\'{\i}so, Chile}
\author{J. A. Helay\"el-Neto}
\email{helayel@cbpf.br}
\affiliation{Centro Brasileiro de Pesquisas F\'\i sicas, Rio de Janeiro, RJ, Brasil}
\author{F. J. L. Leal}
\email{fernando.leal@ifes.edu.br}
\affiliation{Instituto Federal de Educa\c{c}\~ao, Ci\^encia e Tecnologia do Esp\'{\i}rito Santo,
Campus Cariacica, ES, Brasil}
\date{\today }

\begin{abstract}
Background fermion condensates in a landscape dominated by global SUSY are reassessed in connection with a scenario where Lorentz symmetry is violated in the bosonic sector (actually, the photon sector) by a $CPT$-even $k_F$-term. An effective photonic action is discussed that originates from the supersymmetric background fermion condensates. Also, the photino mass emerges in terms of a particular condensate contrary to what happens in the $k_{AF}$-violation. Finally, the interparticle potential induced by the effective photonic action is investigated and a confining profile is identified.
\end{abstract}
\maketitle

\section{Introduction}

Models that realize the breaking of Lorentz symmetry have raised a great deal of interest after Kosteleck\'{y} and Samuel have shown \cite{Samuel}, in a context of bosonic strings , that condensation of tensor fields is dynamically possible, contrary to the physics of the Standard Model , whose dynamics does not yield Lorentz-symmetry violation (LSV). However,  models with LSV are to be considered as effective theories and the analysis of their phenomenological aspects at low energies may provide information and impose constraints on the more fundamental theory from which they stem.

A general framework for testing the low-energy manifestations of CPT-breaking and LSV is the so-called Standard-Model Extension (SME). In this approach, the effective Lagrangian corresponds to the usual Lagrangian of the Standard Model (SM) corrected by SM operators of any dimensionality contracted with suitable Lorentz-violating (LV) tensorial background coefficients. The effective Lagrangian is written in a Lorentz-invariant form so as to ensure what we refer to as observer's independence of the physics of the system under study . However, the physically relevant transformations are those that affect the dynamical variables (fields) that parametrize the system. 
These changes are named  particle transformations, whereas the latter, the coordinate transformations (that include the background tensors) are called observer's transformations. We point out the work of Ref. \cite{Colladay} where these concepts are throughly analyzed.

Concerning the experimental searches for the CPT/LSV, the generality of the SME has provided the basis for many investigations. In the flat spacetime limit, phenomenological  studies include electrons \cite{electron},photons, muons \cite{muon}, mesons \cite{meson}-\cite{meson2}, baryons \cite{barion}, neutrinos\cite{neutrino} and the Higgs \cite{higgs} sector. Gravitational interaction has also been deeply investigated  \cite{gravity}- \cite{gravity2}, \cite{data} and one can set  current limits on the parameters associated to the breaking of relativistic covariance.

The violation of CPT invariance has also been extensively studied in the framework of a modified Dirac theory \cite{Hamilton} and its non-relativistic regime, with the calculation and discussion of the spectrum of the non-relativistic hydrogen atom \cite{Manojr}. In the direction of fermionic models in the presence of LSV, there has been an effort to associate magnetic properties of spinless and/or neutral particles if a non-minimal coupling of the Lorentz-symmetry violating background to fermionic matter and gauge bosons is taken into account (\cite{Phases}, \cite{General}). Still in the realm of atomic physics and optics, we should quote a line of works that set out to examine effects of LSV in electromagnetic cavities and optical systems \cite{Cavity,Masers}, which have finally contributed to set up new bound on the parameters associated to LSV.

The breaking of Lorentz symmetry should be traced back to the dynamics of a
more fundamental physics at  energies much above our present accelerators's energies, for example, at very high energies in astrophysical and even cosmological phenomena. On the other hand, Supersymmetry  (SUSY) should be exact at these energy scales, or, it may happen that it is broken at a scale very close to this primary physical environment. We claim that  LSV and SUSY breakings are not completely independent events in a high-energy regime. We then work with the hypothesis that LSV occurs in a world that is dominated by exact SUSY or still keeps track of a SUSY broken at a slightly higher scale. We highlight the works of Refs. (\cite{LSVSUSY}, \cite{David}, \cite{petrov}), where a list of papers that put SUSY in direct association with models  CPT-breaking and LSV. More recently, the relationship between SUSY breaking and LSV has been discussed in the works by Chkareuli \cite{Chkareuli} and in the article by Pospelov and Tamarit \cite{pos-tam}, where these authors consider  the possibility that SUSY and Lorentz-symmetry breaking have a common origin if supersymmetric matter is coupled to Horava-Lifshitz gravity.

Our proposal here is to place LSV in a scenario where SUSY still holds as an exact symmetry. We shall then notice afterwards that the breaking of Lorentz-symmetry naturally induces SUSY violation, as we shall show in details throughout this paper. With the idea that SUSY is present from the very outset, we claim that the background vector (or tensor) that signals LSV must be component of some particular SUSY, multiplet. This is the key point of our proposal. In a previous paper \cite{susyI}, we have proposed a SUSY-dominated scenario to study LSV by considering the Carroll-Field-Jackiw (CFJ) model, and we have proposed that the background associated to LSV was sitting in a chiral scalar superfield. Our study has revealed that this situation is characterized by a set of fermion condensates that accompany the background vector of the CFJ model. These fermionic pairs turn out to induce physical effects such as mass splitting for supersymmetric partners an a set of extended dispersion relations for the photon and photino sectors. In this direction, we would like to quote the interesting article by Tomboulis \cite{Tomboulis}.

Motivated by the fact that SUSY reveals that LSV is realized with a bosonic background along with a whole set of fermions that condensate in the process, we pursue here another investigation to better understand the issue: we select the so-called $k_F$-term, for which CPT is not broken, and study the effect of the fermion condensates associated to this type of breaking on the physics of the photon and photino. In special, we are very much concerned with the emergence of  an effective photonic action that comes out as a by-product of LSV and the associated fermion condensates. Again, we are going  to conclude  that the  LSV is accompanied by the emergence of a Goldstone fermion, which signals SUSY breaking, even thought no F- or D-term is behind SUSY violation.

The effective photonic model we shall derive carries the fermionic condensates that are in this context messengers of LSV. It may be adopted to reassess the discussion of the emergence of an interparticle potential with a confining piece along with an Yukawa profile whose parameters incorporate the contribution of the bosonic background and fermion condensates. This study reveals that LSV and the supersymmetric dynamics that induce the formation of pairs of fermions may be present in the electrostatic interacting energy of two particles with opposite charges. Our work is organized according to the following structure: Section II is simply the formulation of the component- field action for the supersymmetric version of the $k_F$-term in the case a single four-vector, $\xi_\mu$, is the bosonic signal of LSV. We accommodate $\xi_\mu$ in a chiral scalar superfield and we identify the fermionic condensates that come out in the action with LSV.

Section III is devoted to a simplification of the LSV action by the elimination of auxiliary field present in the gauge potential superfield. In Section IV, we actually start by deriving the physical effects we wish to discuss: photon-photino splitting, dispersion relations an the photon effective action inherited from LSV. Next, in the Section V, the effective photonic action is considered to discuss the electrostatic confining potential between two opposite charges. Finally, we  present our Concluding Comments and future developments in Section VI. Two Appendices follow: in Appendix A, we cast a primary component-field action written in terms of Weyl spinors. Next, in Appendix B, we present a term which is a key algebraic expression for the attainment of the field action that we shall be actually working with throughout our paper.

\section{The $k_F$-term, its reduction and its supersymetric extension }

We start off with the action for the CPT-even term for the abelian gauge sector of Standard Model Extension:

\begin{equation}
S_{\mbox{CPT-even}}=-\frac{1}{4}\int d^4x (k_F)_{\mu\nu\alpha\beta}F^{\mu\nu}F^{\alpha\beta}.
\end{equation}

The tensor $k_F$, from now on written as $K_{\mu\nu\alpha\beta}$ display the properties:
\begin{equation}
K_{\mu\nu\alpha\beta}=-K_{\nu\mu\alpha\beta}=K_{\mu\nu\beta\alpha}=K_{\alpha\beta\mu\nu},
\end{equation}
it is double-traceless and its fully anti-symmetric component is ruled out because it yields a total derivative. As well-known, it depends on 19 parameters.

If moreover we wish to suppress the components that yield birefringence, we end up with only 9 independent components.  According to the ansatz discussed in \cite{ansatz}, we may finally parametrize $K_{\mu\nu\alpha\beta}$ as it follows below:
\begin{equation}
K_{\mu\nu\alpha\beta}=\frac{1}{2}(\eta_{\mu\alpha} {\tilde \kappa}_{\nu\beta}-\eta_{\mu\beta} {\tilde \kappa}_{\nu\alpha}+\eta_{\nu\beta} {\tilde \kappa}_{\mu\alpha}-\eta_{\nu\alpha} {\tilde \kappa}_{\mu\beta}),
\end{equation}
\begin{equation}
{\tilde \kappa}_{\alpha\beta}=(\xi_\alpha \xi_\beta-\eta_{\alpha\beta}\frac{(\xi_\rho \xi^\rho)}{4}),
\end{equation}
and the essence of LSV is traced back to the constant background 4-vector $\xi_\mu$, so that the $k_F$ action becomes 
\begin{equation} 
S=\int d^4x \dfrac{1}{4}\left (\frac{1}{2}\xi_\mu \xi_\nu F^{\mu}_{\,\,\,\kappa}F^{\kappa \nu} +\frac{1}{8}\xi_\rho \xi^\rho F_{\mu\nu}F^{}\mu\nu\right).\label{eq:action}
\end{equation}

In our proposal, this is a more reasonable situation. If we were to identify the whole tensor $K_{\mu\nu\alpha\beta}$ as a component of a given superfield, higher spins (actually, $s= \frac{3}{2}$) would be present in a global SUSY framework. Since we have $\xi_\mu$ as the signal of LSV, no risk of higher fermionic spins in the background is undertaken if the effects of the $K$-tensor are transferred to the $\xi^\mu$-vector.

In the work of  ref. \cite{Belich-Leal}, two ways have been suggested to implement a SUSY-extension for a 4-vector background: $\xi_\mu$ may appear as the gradient of a scalar (in this case, LSV is in a chiral superfield) or a complete vector (with transverse and longitudinal components); in the latter case, $\xi_\mu$ should be a vector component of what we call a vector superfield. To consider a simpler fermionic set partners, we choose to place $\xi^\mu$ in the chiral superfield: in the first case the supersymmetry is implemented through a chiral multiplet and the other by means of an vector multiplet. For simplicity, we work only the chiral case . In this proposal, the extended action written in superfield formalism is:
\begin{equation}  
S^{(susy)}_{\mbox{CPT-even}}=\int d^4x d^2 \theta d^2 \bar{\theta}\,\, \Big{[}(D^\alpha \Omega) W_{\alpha} (\bar{D}_{\dot{\alpha}} \bar{ \Omega})\bar{ W}^{\dot{\alpha}}+h.c\,\Big{]} =S_{ferm}+S_{boson}+S_{mixing}, \label{cbpf05}
\end{equation}
where
\begin{eqnarray}
W_\alpha(x)=&&\lambda_\alpha(x)+i\theta \sigma^\mu \bar{\theta} \partial_\mu \lambda_\alpha(x)-\frac{1}{4}\bar{\theta}^2\theta^2\Box \lambda_\alpha(x)+
2 \theta_\alpha D(x)-i\theta^2 (\bar{\theta}\sigma^\mu)_\alpha \partial_\mu D(x)+(\sigma^{\mu\nu}\theta)_\alpha F_{\mu\nu}(x)\notag\\
&&-\frac{1}{2}\theta^2 (\sigma^{\mu\nu} \sigma^{\rho} )_\alpha \partial_\rho F_{\mu\nu}(x)-i(\sigma^\mu \partial_\mu \lambda[x])_\alpha\theta^2
\end{eqnarray}

is the well-known field-strength superfield  ($\lambda$ is the photino, $ F_{\mu\nu}$ the usual gauge-field strength and $D$ the auxiliary field); the chiral background superfield, $\Omega$,  is $\theta$-expanded as follows
\begin{eqnarray}
\Omega(x)=&&S(x)+\sqrt{2}\theta \zeta(x)+i\theta \sigma^\mu\bar{\theta}\partial_\mu S(x)+
\theta^2 G(x)+\frac{i}{\sqrt{2}}\theta^2 \bar{\theta}\bar{\sigma}^\mu \partial_\mu \zeta(x)-\frac{1}{4}\bar{\theta}^2 \theta^2 \Box S(x),
\end{eqnarray}
where $S$ and $G$ are complex scalars and $\zeta$ is a Weyl component of a Majorana fermion. By projecting the action (6) into component fields, we readily get that $\xi_\mu=\partial_\mu S$ and the $S_{boson}$, $S_{ferm}$ and $S_{mixing}$ may be found, in terms of Weyl spinors in Appendix A. We prefer to quote below the component-field action directly in terms of Majorana spinors, for it is much simpler and one can control much more easily the various couplings present in the action.

At this point, we also make a special consideration about the background superfield $ \Omega$: taking $S$ linear in $x^\mu$ ($S=\xi_\mu x^\mu$, $\xi_\mu$ constant), $\partial_\mu \zeta=0$ and $G=0$, is compatible with SUSY, in the sense that these properties are kept if global SUSY transformations are done, and moreover we reproduce the $k_F$-term as we wish from the very beginning. 
Now, we shall move on with two purposes:
\begin{itemize}
\item (i) to rewrite the whole action in terms of 4-components Majorana spinors, $Z\equiv(\zeta \,\,\,\bar{\zeta})^t$ and $\Lambda \equiv(\lambda\,\,\, \bar{\lambda})^t$,
\item (ii) to Fiery-Rearrange the terms in $S_{ferm}$ where the fermions $\zeta$ and $\lambda$ are mixed. This process selects for us 3 types of background  fermion condensates (already written in terms of Majorana spinors):
\begin{eqnarray}
\theta=&&\bar{Z}Z\notag\\
\tau=&&\bar{Z}\gamma_5 Z\notag\\
C^{\mu}=&&\bar{Z}\gamma^\mu \gamma_5 Z,
\end{eqnarray}
\end{itemize}
for which the relations below hold true:
\begin{equation}
\theta^2=-\tau^2=\frac{1}{4}C^\mu C_\mu\notag,
\end{equation}
\begin{equation}
\theta \tau=\theta C^\mu= \tau C^\mu=0.
\end{equation}

With all these considerations, the action (\ref{cbpf05}) can be brought into a more readable form:
\begin{subequations}
\begin{align}
S_{boson}= \int d^4x \Big[ &D^2 (32|G|^2+16\partial_\mu S \partial^\mu S^*)+8iD F^{\mu\nu}(\partial_\mu S\partial_\nu S^*- \partial_\mu S^*\partial_\nu S) \nonumber\\
&-8F^{\mu\kappa}F_{\kappa}^{\,\,\,\,\nu}(\partial_\mu S\partial_\nu S^*+ \partial_\mu S^*\partial_\nu S)-4F^{\mu\nu}F_{\mu\nu} \partial_\alpha S \partial_\alpha S^* \,\,\,\Big]; \label{val05}\\
S_{ferm}=& \int dx^4 (C^\mu \bar{ \Lambda}   \gamma^\nu \gamma_5\partial_\mu\partial_\nu \Lambda+qC_\mu \bar{ \Lambda}   \gamma^\mu\gamma_5\Box \Lambda),\,\,\,\,\,\,\,\mbox{where q is a numerical factor}\,\,\,\, (q=\frac{4-\sqrt{2}}{16}); \label{val10}\\
S_{mixing}=& \int d^4x \Big[  D \Big{(}10\sqrt{2}Re(\partial_\mu S)(\bar{Z}\partial_\mu \Lambda)-8\sqrt{2}i Re(\partial_\mu S) (\bar{Z}\Sigma^{\mu\nu}\partial_\nu \Lambda)\notag\\
&8\sqrt{2} Im(\partial_\mu S)(\bar{Z}\Sigma^{\mu\nu}\gamma_5 \partial_\nu \Lambda)+10\sqrt{2}i Im(\partial_\mu S)(\bar{Z}\gamma_5 \partial^\mu \Lambda)\Big{)}\notag\\
&-3\sqrt{2}Im(\partial_\nu S)[\partial_\mu F^{\mu\nu}]\bar{Z} \Lambda+3\sqrt{2}Re(\partial_\nu S)[\partial_\mu F^{\mu\nu}]\bar{Z}\gamma_5 \Lambda+\notag\\
&4\sqrt{2}i \partial_{[\nu}F_{\mu]\alpha} Im(\partial^{\alpha}S)\bar{Z}\Sigma^{\mu\nu}\Lambda +4\sqrt{2}i \partial_{[\nu}\tilde{F}_{\mu]\alpha} Re(\partial^{\alpha}S)\bar{Z}\Sigma^{\mu\nu}\Lambda \notag+\\
&4\sqrt{2} \partial_{[\nu}F_{\mu]\alpha} Im(\partial^{\alpha}S)\bar{Z}\Sigma^{\mu\nu}\gamma_5\Lambda +4\sqrt{2} \partial_{[\nu}\tilde{F}_{\mu]\alpha} Re(\partial^{\alpha}S)\bar{Z}\Sigma^{\mu\nu}\gamma_5\Lambda\Big]. \label{val15}
\end{align}
\end{subequations}

$D$ appears as an auxiliary field and we are going, in the next Section, to eliminate it upon use of its corresponding equation of motion.

\section{Eliminating the  auxiliary field }

The equations above are indeed more manageable to work. In order to complete our model, we must add up to equations ($11$) the supersymmetric version of the Maxwell action.  After this is done, its advisable to eliminate  eliminate the auxiliary field, $D$, by means of algebraic equation of motion. Notice that the total action can be written  in terms of auxiliary field in the form below: 
\begin{equation}
S^{(full)}=S^{(susy)}_{\mbox{Maxwell}}+S^{(susy)}_{\mbox{CPT-even}}=S+\int d^4x \,\beta D +\int d^4x \,\alpha D^2,
\end{equation}
\begin{equation}
S^{(full)}=S-\int dx^4 \,\frac{\beta^2}{2(2+\alpha)},
\end{equation}
where $\alpha$ and $\beta$ are expressed in terms of background and fields in the gauge sector as follows: 
\begin{eqnarray}
\alpha=&&16(\partial_\kappa S \partial^\kappa S^*),\notag\\
\beta{}=&&10\sqrt{2} Re(\partial_\mu S)(\bar{Z}\partial_\mu \Lambda)-8\sqrt{2}i Re(\partial_\mu S) (\bar{Z}\Sigma^{\mu\nu}\partial_\nu \Lambda) + \notag\\
&&+ 8\sqrt{2}  Im(\partial_\mu S)(\bar{Z}\Sigma^{\mu\nu}\gamma_5 \partial_\nu \Lambda)+10\sqrt{2}i Im(\partial_\mu S)(\bar{Z}\gamma_5 \partial^\mu \Lambda)+16\,m_{\mu\nu}F^{\mu\nu}, \label{cbpf15}
\end{eqnarray}
where $m_{\mu\nu}=Re(\partial_\mu S)Im(\partial_\nu S)-Re(\partial_\nu S)Im(\partial_\mu S)$.\\

The calculation of $\beta^2$ involves again the use of Fierz identities and properties of bilinear formed by anticommuting Majorana spinors. The final result is somehow cumbersome, so that, to keep the balance of the text, we believe it is advisable to collect the result in an Appendix. For that, we have included the Appendix B.

Then, incorporating the $\beta^2$-term in to the action, we have:
\begin{eqnarray}
S^{\mbox{(full)}}=\int dx^4{\Big{[}} &&-\frac{1}{4}F^{\mu\nu}F_{\mu\nu}+K_{\mu\nu\alpha\beta}F^{\mu\nu}F^{\alpha\beta}-\dfrac{64}{(1+8\partial_\rho S \partial^\rho S^*)}m_{\mu\nu}m_{\alpha\beta}F^{\mu\nu}F^{\alpha\beta}+\notag\\
&&-\bar{\Lambda}\dfrac{\tilde{a}}{4(1+8 \partial_\kappa S \partial^\kappa S^*)}\Lambda-\bar{\Lambda}\dfrac{\tilde{b}}{4(1+8 \partial_\kappa S \partial^\kappa S^*)}\gamma_5\Lambda+\notag\\
&&\bar{\Lambda}\Big{[}-(C,\partial)\partial_\mu+q  \Box C_\mu+C^\alpha d_{\alpha\mu}\Big{]}\gamma^\mu\gamma_5 \Lambda+2 \bar{Z}N \Lambda \,\,\,{\Big{]}}.
\end{eqnarray}
The coefficients $m_{\mu\nu}$ and $d_{\alpha\mu}$, ${\tilde{a}}$ and ${\tilde{b}}$ can all be found in Appendix B. The $N$-matrix above, that mixes the background fermion and the photino, is given by a lengthly expression that involves the photon field  its field strength, $F_{\mu\nu}$. This term mixes therefore the photon and the photino fields, and the explicit form of $N$ follows below:
\begin{subequations}
\begin{align}
N=&I^{(1)}+iI^{(2)}\gamma_5+iI_{\mu\nu}\Sigma^{\mu\nu},\\
&\mbox{where}\notag\\
I^{(1)}=&-\frac{3}{2}\sqrt{2} Im(\partial_\mu S) \partial_\alpha F^{\alpha\mu}+\dfrac{20\sqrt{2}}{(1+8 \partial_\kappa S \partial^\kappa S^*)}m_{\alpha\beta}Re(\partial_\rho S) \partial^\rho F^{\alpha\beta},\\
I^{(2)}=&\frac{3}{2}\sqrt{2} Re(\partial_\mu S) \partial_\alpha F^{\alpha\mu}+\dfrac{20\sqrt{2}i}{(1+8 \partial_\kappa S \partial^\kappa S^*)}m_{\alpha\beta}Im(\partial_\rho S) \partial^\rho F^{\alpha\beta},\\
I_{\mu\nu}=&2\sqrt{2}\Big{[} Im(\partial^\alpha S) \partial_{[\nu}F_{\mu]\alpha}+Re(\partial^\alpha S) \partial_{[\nu}\tilde{F}_{\mu]\alpha}\Big{]}-\dfrac{16\sqrt{2}}{(1+8 \partial_\kappa S \partial^\kappa S^*)}m_{\alpha\beta}Re(\partial_\mu S) \partial_\nu F^{\alpha\beta}\notag\\
&\sqrt{2} \epsilon_{\alpha\beta\mu\nu}\Big{[} Re(\partial_{\rho} S )\partial^\alpha \tilde{F}^{\beta\rho}-Im (\partial_{\rho} S ) \partial^\alpha\tilde{F}^{\beta\rho}\Big{]}-\dfrac{8\sqrt{2}}{(1+8 \partial_\kappa S \partial^\kappa S^*)}\epsilon_{\alpha\beta\mu\nu}m_{\theta\lambda}Re(\partial^\alpha S) \partial^\beta F^{\theta\lambda}.
\end{align}
\end{subequations}

Let call the reader's attention to the fact that the $A^\mu-\Lambda$  mixed term appears in the form $\bar{Z} N \Lambda$; the $N$-matrix is written in terms of $1$, $\gamma_5$ and $\Sigma_{\mu\nu}$, and the coefficients $I^{(1)}$, $I^{(2)}$ and $I_{{\mu\nu}}$ contain terms in the background field $S$ (through $\partial_\mu S$) and $F_{\mu\nu}$. As a whole, the term $\bar{Z} N \Lambda$ is quadratic in the bosonic background and quadratic (but non-diagonal) in the degrees of freedom of the gauge sector ($A^\mu$ and $\Lambda$).

\section{ Dispersion relations  and A Purely  Photonic Efective  Action}

The $N$-matrix previously defined  depends on the field strength, $F^{\alpha\beta}$, through terms of the form $\partial^\mu F^{\alpha\beta}$. Let us then, for convenience, introduce the following form for  carry the field strength $F^{\mu\nu}$, we introduce the following form for $N=N^{'}_{\alpha}A^{\alpha}$. This allows us  to  rewrite in a more compact form the quadratic action in the photon and photino fields. We unify  the latter in a sort of doublet: $ \Psi\equiv (\begin{array}{c}
\Lambda\\
A_\nu \end{array})$, $\bar{\Psi}\equiv (\bar{\Lambda} \,\,\,A_\mu)$, so that the full action may be thought into the form
\begin{equation}
S^{\mbox{(full)}}=\frac{1}{2}\int dx^4 \bar{\Psi} {\cal{O}} \Psi,\\
\end{equation}
where the matrix operator ${\cal O}$ is given by 
\begin{equation}  {\cal{O}}= \left( \begin{array}{cc}
M &\,\,\,N^{'}\\
N^{'}& \,\,\,Q\\
\end{array} \right),
\end{equation}
with the sub-matrices given as below:
\begin{subequations}
\begin{align}
M=&-\dfrac{\tilde{a}}{4(1+8 \partial_\kappa S \partial^\kappa S^*)}1_{4\times 4}-\dfrac{\tilde{b}}{4(1+8 \partial_\kappa S \partial^\kappa S^*)}\gamma_5+\notag\\
&-\dfrac{p^\mu}{2}\gamma_\mu+\Big{(}(p,C)p_\mu-q  p^2 C_\mu+C^\alpha d_{\alpha\mu}\Big{)}\gamma^\mu\gamma_5,\\
Q_{\mu\nu}=&-\frac{1}{2}\Box\,\theta_{\mu\nu}+(J_{\mu\alpha\beta\nu}-J_{\mu\alpha\nu\beta}+J_{\alpha\mu\nu\beta}-J_{\alpha\mu\beta\nu})\Box\, \omega^{\alpha\beta}\notag,\\
where
&\notag\\
J_{\mu\alpha\beta\nu}=&K_{\mu\alpha\beta\nu}-\dfrac{64}{(1+8\partial_\rho S \partial^\rho S^*)}m_{\mu\alpha}m_{\beta\nu}.
\end{align}
\end{subequations}

Recalling the expression for $m_{\mu\nu}$ in the set of eqs. (\ref{cbpf15}) and that, without loss of generality, we are taking the (complex) scalar background, $S$, linear in $x^\mu$  ($S = {\xi _\mu }{x^\mu }$), we are safely allowed to consider $\xi^\mu$ as real, which yields a vanishing $m_{\mu\nu}$. However, should we take $\xi^\mu$ as a constant complex $4$-vector, $m_{\mu\nu}$ simply becomes a constant and this constant does not introduce any physical effect that we miss once we adopt $\xi^\mu$ to be real.

We recall  that  the elimination of auxiliary field, $D$, yields a  new contribution to the usual Lorentz-breaking tensor, $K_{\mu\nu\alpha\beta}$,  which is shown in the tensor $J_{\mu\nu\alpha\beta}$. A conventional procedure would consist in explicitly  calculating ${\cal O}^{-1}$ in order to get the propagators  $\bar{\Lambda} \Lambda-$, $\Lambda A_\mu-$ and  $A_\mu A_\nu-$ propagators, whose pole structure corresponds to the dispersion relation. However, if we are simply interested in the dispersion relations   for the photon and photino fields, we can concentrate only on  the matrices  $M$ and $Q$, as was have shown in more details in the  paper of  Ref.\cite{susyI}. Actually, the poles of  the photon and photino propagators can be read off from  $\mbox{det} Q=0$ and $\mbox{det} N =0$, respectively.

The photino propagator corresponds to the inverse matrix $M^{-1}$,  whose pole structure is found in det $M$:
\begin{equation}
M^{-1}=A+B\gamma_5+v_\theta \gamma^\theta+\omega_\theta \gamma^\theta\gamma_5,
\end{equation}
whith the  coefficients given  by:
\begin{equation}
A=\dfrac{\tilde{a}p^2}{16(1+8 \partial_\kappa S \partial^\kappa S^*)\Delta},
\end{equation}
\begin{equation}
B=-\dfrac{\tilde{b}p^2}{16(1+8 \partial_\kappa S \partial^\kappa S^*)\Delta},
\end{equation}
\begin{equation}
v_\mu=\Big{[}\dfrac{\tilde{a}^2-\tilde{b}^2}{16(1+8 \partial_\kappa S \partial^\kappa S^*)^2}-\frac{p^2}{4}-w^2 \Big{]}\frac{p_\mu}{2\Delta}+\dfrac{(w,p)w_\mu}{\Delta},
\end{equation}
\begin{equation}
\omega_\mu=(1-q)p^2(p,C)\dfrac{p_\mu}{2\Delta}+(C^\alpha p^\beta d_{\alpha\beta})\dfrac{p_\mu}{2\Delta}-\dfrac{p^2}{4\Delta}\Big{[} (p,C)p_\mu-q  p^2 C_\mu+C^\alpha d_{\alpha\mu}\Big{]},
\end{equation}
and 
\begin{equation}
\Delta=\frac{p^4}{16}-(p,\tilde{w})^2-\dfrac{p^2\tilde{a}^2}{32(1+8k \partial_\kappa S \partial^\kappa S^*)^2}+\dfrac{p^2\tilde{b}^2}{32(1+8k \partial_\kappa S \partial^\kappa S^*)^2}+\frac{p^2}{2}\tilde{w}^2.
\end{equation}

We can separate the denominator $\Delta$ in two parts: one containing terms up to 2nd. order in powers of $\partial_\mu S$ and the another piece that only contains higher powers in $\partial_\mu S$ . This splitting is suitable if we recall that the LSV parameters are very tiny, so that we confine our considerations to terms which are second order in $\partial_\mu S$, and we collect higher terms in ${\cal{O}}(3)$:
\begin{equation}
\Delta=p^4\theta^2\tilde{\Delta}=p^4\theta^2\Big{(}\frac{1}{16\theta^2}+\left[ C^{(1)}p^2+C^{(2)}_{\mu\nu}p^\mu p^\nu \right]+{\cal O}(3)\Big{)}, \label{sp05}
\end{equation}
where
\begin{equation}
C^{(1)}=(q^2-q-\frac{1}{2})+\left[\dfrac{1}{(1+8\partial_\mu S \partial^\mu S^*)} \right](4q-2)(\eta_{\mu\nu}t^{\mu\nu}),  \label{sp10}
\end{equation}
\begin{equation}
C^{(2)}_{\mu\nu}=\left[\dfrac{1}{2(1+8\partial_\mu S \partial^\mu S^*)} \right][42q-29]t_{\mu\nu}. \label{sp15}
\end{equation}

Since $K_{\mu\nu\alpha\beta}$ is a linear combination of bilinear in $\partial_\mu S$, terms of ${\cal{O}}(3)$ or higher in eq. (15) are dropped out . We also notice that the coefficient $C^{(2)}_{\mu\nu}$ is much smaller than $C^{(1)}$ since $|t_{\mu\nu}|<< 1$, so, in this approximation, it is possible remove the term that mixes the momenta and we find a very simple dispersion relation for the photino which is given by 
\begin{equation}
\Delta^{(\mbox{aprox})}= C^{(1)}\theta^2 p^4(p^2-m^2)=0, \label{leblon}
\end{equation}
with 
\begin{equation}
m_{photino}^2 =  - \frac{{{C^{\left( 1 \right)}}}}{{16{\theta ^2}}}, \label{sp20}
\end{equation}
notice that $C^{(1)}$ is negative. Here, contrary to the Carrol-Field-Jackiw supersymmetrised model of Ref. \cite{susyI}, the photino mass carries an explicit dependence on the $\theta$-fermion condensate. This is a new feature of the $k_F$-model.

Following along analogous steps, we are able to find the dispersion relation for the photon
\begin{equation}
p^0_{\pm}=(1+\rho \pm \sigma)|\bar{p}|, \label{sp25}
\end{equation}
where $\rho=\frac{1}{2}\tilde{K}_{\alpha}^{\,\,\,\alpha}$ and $\sigma^2=\frac{1}{2}(\tilde{K}_{\alpha\beta})^2-\rho^2$, with $\tilde{K}^{\alpha\beta}=K^{\alpha\beta\mu\nu}\hat{p}_\mu \hat{p}_\nu$ and $\hat{p}^\mu=p^\mu /|\bar{p}|$.\\

Finally, by eliminating the mixed $A_\mu \, \Lambda$ terms,  we shall find an effective action for purely photonic sector  . In the action, the term that combine  these fields is given by $2 \bar{Z}N \Lambda$. We notice that this term can be removed in we perform a convenient shift in the photino field. By redefining the fermion field according to  $\Upsilon=\Lambda+M^{-1}\bar{N}Z$, we attain a new action that is totally diagonal in the fields $\Upsilon$ and $A_\mu$.
With the help of the properties of the fermionic condensates (7) and the gamma-matrix algebra, the redefinition of $\Lambda$ suggested above yields an effective term for the photon sector which can be expressed as follows:
\begin{eqnarray}
S_{\mbox{effective}}^{\,(photon)}=&&\int d^4 x  \bar{Z}(N M^{-1} \bar{N})Z\notag\\
=&&\int d^4 x\Big{[} (I^{(1)}I^{(1)}-I^{(2)}I^{(2)}+\frac{1}{2}I_{\mu\nu}I^{\mu\nu})(A\theta+B\tau)+i(2I^{(1)}I^{(2)}-\frac{1}{2}I_{\mu\nu}\tilde{I}^{\mu\nu})(A\tau+B \theta)+\notag\\
&&(I^{(1)}I^{(1)}+I^{(2)}I^{(2)}+\frac{1}{2}I_{\mu\nu}I^{\mu\nu})\omega_\theta C^\theta+2I^{(1)}I^{\theta\rho}\omega_\theta C_{\rho} -2I^{(2)}\tilde{I}^{\theta\rho}\omega_\theta C_{\rho}\Big{]},
\end{eqnarray}
where the coefficients $I^{(1)}, I^{(2)}$ and $I_{\mu\nu}$ only exhibit derivatives of the field strength.

Taking into account the previous discussion on the approximation we adopt to treat the LSV parameters, we can also ignore the terms of order ${\cal O}(3)$ in Eq.(16), so that the full effective Lagrangian density  (in the momentum space) for the photon is given by the expression
\begin{equation}
{\cal L}={\cal L}_{\mbox{old}}+{\cal L}_{\mbox{effective}}, \label{effe05}
\end{equation}
where
\begin{equation}
{\cal L}_{\mbox{old}}=-\frac{1}{4}F_{\mu\nu}F^{\mu\nu}-16 t_{\mu\nu}F^{\mu\kappa}F_\kappa^{\,\,\,\nu}-4 F_{\mu\nu}F^{\mu\nu}(t_{\alpha\beta}\eta^{\alpha\beta}), \label{effe10}
\end{equation}
and
\begin{equation}
{\cal L}_{\mbox{effective}}=\dfrac{1}{\tilde{\Delta}}(\frac{q}{4}-\frac{1}{8})t_{\rho\lambda}\big[4p^2 F_{\mu}^{\,\,\,\,\rho}F^{\mu\lambda}-\eta^{\rho\lambda} p^2 F_{\mu\nu}F^{\mu\nu} \big]+
\dfrac{1 }{\tilde{\Delta}}(\frac{5q}{8}+\frac{13}{16})t_{\rho\lambda} \big[ p_\mu p_\nu F^{\mu\rho} F^{\nu\lambda}\big].\label{effe15}
\end{equation}
So that, in the supersymmetric scenario for the $k_F-$Lorentz symmetry breaking, the purely (effective) photonic action is completely given by $\partial F$-terms. This is to be compared with the corresponding effective photonic action worked out in \cite{susyI}, where only $FF$-terms have shown up. We recall that $t_{\mu\nu}$ may be found in Appendix B.

\section{Interaction energy}

We now examine the interaction energy from the viewpoint of the gauge-invariant but path-dependent variables formalism, along the lines of Refs.\cite{GaeteSch, GaeteEuro1, GaeteEuro2, Gaete99, susyI}. This can be done by computing the expectation value of the energy operator $H$ in the physical state $|\Phi\rangle$ describing the sources, which we will denote by ${\langle H \rangle}_{\Phi}$. The starting point is the effective Lagrangian density:
\begin{eqnarray}
{\cal L} &=&  - \frac{1}{4}F_{\mu \nu } \left[ {1 + 16t_\alpha ^\alpha   - 4\left( {\frac{q}{4} - \frac{1}{8}} \right)t_\alpha ^\alpha  \frac{\Delta }{{\tilde \Delta }}} \right]F^{\mu \nu }  -16t_{\mu \nu } F^{\mu \lambda } F^{\nu}\ _{\lambda}     - 4\left( {\frac{q}{4} - \frac{1}{8}} \right)t_{\rho \lambda } F^{\mu \lambda } \frac{\Delta }{{\tilde \Delta }}F_\mu ^\rho   \nonumber\\ 
&-& \left( {\frac{{5q}}{8} + \frac{{13}}{{16}}} \right)t_{\rho \lambda } F^{\mu \lambda } \frac{{\partial _\mu  \partial _\nu  }}{{\tilde \Delta }}F^{\nu \rho }, \label{susyII-05} 
\end{eqnarray}
where $\Delta  \equiv \partial _\mu  \partial ^\mu$. However, as was mentioned before, this paper is aimed at studying the static potential of the above theory, a consequence of this is that one may replace $\Delta$ by $-\nabla ^2$ in Eq.(\ref{susyII-05}). Furthermore, we recall that the only non-vanishing $t_{\mu \nu }$-terms are the diagonal ones, since, as already anticipated, we can take ${\xi ^\mu }$ real, and, as a symmetric matrix, $t_{\mu \nu }$ can be brought into a diagonal form. Without loss of generality, we may always choose $t_{00}\ne 0$. 

Therefore, the effective Lagrangian becomes 
\begin{equation}
{\cal L} =  - \frac{1}{4}\gamma F_{\mu \nu } \frac{{\left( {\nabla ^2  - M^2 } \right)}}{{\left( {\nabla ^2  - m^2 } \right)}}F^{\mu \nu }  +16t_{0 0 } F_{i 0 } F^{i0}   - \frac{{A_4 }}{{A_2 }}t_{00} F_{i 0 } \frac{{\nabla ^2 }}{{\left( {\nabla ^2  - m^2 } \right)}}F^{i 0}   - \frac{{A_5 }}{{A_2 }}t_{00} F^{i 0} \frac{{\partial _i  \partial _j  }}{{\left( {\nabla ^2  - m^2 } \right)}}F^{j 0}.\label{susyII-10} 
\end{equation}
Here, $\gamma  = \frac{{A_3 A_2  - A_4 t_\alpha ^\alpha  }}{{A_2 }}$, $M^2  = \frac{{A_3 A_1 }}{{A_3 A_2  - A_4 t_\alpha ^\alpha  }}$, $m^2  = \frac{{A_1 }}{{A_2 }}$. Whereas that   $ A_1  = \frac{1}{{16\theta ^2 }}$, $ A_2  = k^ \prime   - C^{\left( 1 \right)}$, $A_3  = \left( {1 + 16t_\alpha ^\alpha  } \right)$, $ A_4  = 4\left( {\frac{q}{4} - \frac{1}{8}} \right)$ and $A_5  = \left( {\frac{{5q}}{8} + \frac{{13}}{{16}}} \right)$.\\                     

To obtain the corresponding Hamiltonian, the canonical quantization of this theory from the Hamiltonian point of view is straightforward. The canonical momenta are found to be $\Pi^0=0$ and $\Pi ^i  = \frac{{\left( {\alpha \nabla ^2  - \beta } \right)}}{{\left( {\nabla ^2  - m^2 } \right)}}F^{i0}  + \frac{{2A_5 }}{{A_2 }}t_{00} \frac{{\partial ^i \partial _j }}{{\left( {\nabla ^2  - m^2 } \right)}}F^{j0}$, where $\alpha  = \gamma  -32t_{00}  + 2\frac{{A_4 }}{{A_2 }}$ and $\beta  = \gamma M^2  -32
t_{00} m^2$. Thus, the canonical Hamiltonian takes the form 
\begin{eqnarray}
H_C &=& \int {d^3 x}\Biggl[- A_0 \partial _i \Pi ^i 
-\frac{1}{2}\Pi ^i \frac{{\left( {\nabla ^2  - m^2 } \right)}}{{\left( {\alpha \nabla ^2  - \beta } \right)}}\Pi _i  + \frac{1}{4}F_{ij}\left( {\frac{{\nabla ^2 - M^2}}{{\nabla ^2 -m^2 }}} \right)F^{ij}+
\frac{1}{2}\partial ^i \partial ^k \Pi ^k \frac{{\left( {\nabla ^2  - m^2 } \right)}}{{\left( {\nabla ^2  + \Omega ^2 } \right)^2 \left( {\alpha \nabla ^2  - \beta } \right)}}\partial _i \partial _j \Pi _j \notag \\
&+&
\frac{{A_5 }}{{A_2 }}t_{00} \left( {\frac{{\left( {\nabla ^2  - m^2 } \right)}}{{\left( {\alpha \nabla ^2  - \beta } \right)}}\Pi _i  + \frac{{\left( {\nabla ^2  - m^2 } \right)\partial _i \partial _k \Pi _k }}{{\left( {\nabla ^2  + \Omega ^2 } \right)\left( {\alpha \nabla ^2  - \beta } \right)}}} \right)\frac{{\partial _i \partial _j }}{{\left( {\nabla ^2  - m^2 } \right)}}\left( {\frac{{\left( {\nabla ^2  - m^2 } \right)}}{{\left( {\alpha \nabla ^2  - \beta } \right)}}\Pi ^j  + \frac{{\left( {\nabla ^2  - m^2 } \right)\partial ^j \partial ^m \Pi ^m }}{{\left( {\nabla ^2  + \Omega ^2 } \right)\left( {\alpha \nabla ^2  - \beta } \right)}}} \right)\Biggr]\ . \notag \\
\label{susyII-15} 
\end{eqnarray}

Time conservation of the primary constraint, $\Pi_0=0$, leads to the usual Gauss constraint $\Gamma_1\equiv \partial _i \Pi ^i =0$. The preservation of $\Gamma_1$ for all times does not give rise to any further constraints. The extended
Hamiltonian that generates translations in time then reads $H = H_C + \int {d^3 } x\left( {c_0 \left( x \right)\Pi _0 \left( x \right) + c_1 \left( x\right)\Gamma _1 \left( x \right)} \right)$, where $c_0 \left( x \right)$ and $c_1 \left( x \right)$ are the Lagrange multiplier fields. Since $\Pi^0 = 0$ for all time and $\dot{A}_0 \left( x \right)= \left[ {A_0\left( x \right),H} \right] = c_0 \left( x \right)$ which is completely arbitrary, we discard $A^0 $ nor $\Pi^0$ because they add nothing to the description of the system. The extended Hamiltonian then becomes 
\begin{eqnarray}
H &=& \int {d^3 x}\Biggl[c(x) \left( {\partial _i \Pi ^i} \right) -\frac{1}{2}\Pi ^i \frac{{\left( {\nabla ^2  - m^2 } \right)}}{{\left( {\alpha \nabla ^2  - \beta } \right)}}\Pi _i  + \frac{1}{4}F_{ij}\left( {\frac{{\nabla ^2 - M^2}}{{\nabla ^2 -m^2 }}} \right)F^{ij}+ \frac{1}{2}\partial ^i \partial ^k \Pi ^k \frac{{\left( {\nabla ^2  - m^2 } \right)}}{{\left( {\nabla ^2  + \Omega ^2 } \right)^2 \left( {\alpha \nabla ^2  - \beta } \right)}}\partial _i \partial _j \Pi _j \notag \\
&+&
\frac{{A_5 }}{{A_2 }}a_{00} \left( {\frac{{\left( {\nabla ^2  - m^2 } \right)}}{{\left( {\alpha \nabla ^2  - \beta } \right)}}\Pi _i  + \frac{{\left( {\nabla ^2  - m^2 } \right)\partial _i \partial _k \Pi _k }}{{\left( {\nabla ^2  + \Omega ^2 } \right)\left( {\alpha \nabla ^2  - \beta } \right)}}} \right)\frac{{\partial _i \partial _j }}{{\left( {\nabla ^2  - m^2 } \right)}}\left( {\frac{{\left( {\nabla ^2  - m^2 } \right)}}{{\left( {\alpha \nabla ^2  - \beta } \right)}}\Pi ^j  + \frac{{\left( {\nabla ^2  - m^2 } \right)\partial ^j \partial ^m \Pi ^m }}{{\left( {\nabla ^2  + \Omega ^2 } \right)\left( {\alpha \nabla ^2  - \beta } \right)}}} \right)\Biggr]\ , \notag \\\label{susyII-20} 
\end{eqnarray}
where $c(x) = c_1 (x) - A_0 (x)$.

In order to fix gauge symmetry we adopt the gauge discussed in our previous works, that is,

\begin{equation}
\Gamma _2 \left( x \right) \equiv \int\limits_{C_{\zeta x} } {dz^\nu } A_\nu\left( z \right) \equiv \int\limits_0^1 {d\lambda x^i } A_i \left( {\lambda x } \right) = 0,  \label{susyII-25}
\end{equation}
where $\lambda$ $(0\leq \lambda\leq1)$ is the parameter describing the space-like straight path $x^i = \zeta ^i + \lambda \left( {x - \zeta}\right)^i $ , and $\zeta $ is a fixed point (reference point). There is no essential loss of generality if we restrict our considerations to $\zeta^i=0 $. In this case, the only non-vanishing equal-time Dirac bracket is 

\begin{equation}
\left\{ {A_i \left( x \right),\Pi ^j \left( y \right)} \right\}^ * =\delta{\
_i^j} \delta ^{\left( 3 \right)} \left( {x - y} \right) - \partial _i^x
\int\limits_0^1 {d\lambda x^j } \delta ^{\left( 3 \right)} \left( {\lambda x- y} \right).  \label{susyII-30}
\end{equation}

We now turn to the problem of obtaining the interaction energy between point-like sources in the model under consideration, where a fermion is localized at $\mathbf{y}\prime$ and an antifermion at $\mathbf{y}$. One 
might show that the interaction energy is
\begin{eqnarray}
\left\langle H \right\rangle _\Phi &=& \left\langle \Phi \right|\int {d^3 x} 
\Biggl[-\frac{1}{2}\Pi ^i \frac{{\left( {\nabla ^2  - m^2 } \right)}}{{\left( {\alpha \nabla ^2  - \beta } \right)}}\Pi _i  + \frac{1}{4}F_{ij}\left( {\frac{{\nabla ^2 - M^2}}{{\nabla ^2 -m^2 }}} \right)F^{ij}\Biggr]|\Phi\rangle .  \label{susyll-35}
\end{eqnarray}

Next, the physical state is constructed, following Dirac \cite{Dirac}, 
\begin{equation}
\left| \Phi \right\rangle \equiv \left| {\overline \Psi \left( \mathbf{y }
\right)\Psi \left( \mathbf{y}\prime \right)} \right\rangle = \overline \psi
\left( \mathbf{y }\right)\exp \left( {iq\int\limits_{\mathbf{y}\prime}^{ 
\mathbf{y}} {dz^i } A_i \left( z \right)} \right)\psi \left(\mathbf{y}\prime
\right)\left| 0 \right\rangle,  \label{susyII-40}
\end{equation}
where $\left| 0 \right\rangle$ is the physical vacuum state and the line
integral appearing in the above expression is along a space-like path
starting at $\mathbf{y}\prime$ and ending at $\mathbf{y}$, on a fixed time
slice. 

From the foregoing Hamiltonian structure we then easily verify that 

\begin{equation}
\Pi _i \left( x \right)\left| {\overline \Psi \left( \mathbf{y }\right)\Psi
\left( {\mathbf{y}^ \prime } \right)} \right\rangle = \overline \Psi \left( 
\mathbf{y }\right)\Psi \left( {\mathbf{y}^ \prime } \right)\Pi _i \left( x
\right)\left| 0 \right\rangle + q\int_ {\mathbf{y}}^{\mathbf{y}^ \prime } {\
dz_i \delta ^{\left( 3 \right)} \left( \mathbf{z - x} \right)} \left| \Phi
\right\rangle.  \label{susyII-45}
\end{equation}

In such a case $\left\langle H \right\rangle _\Phi$ reduces to 
\begin{equation}
\left\langle H \right\rangle _\Phi = \left\langle H \right\rangle _0 +
\left\langle H \right\rangle _\Phi ^{\left( 1 \right)} + \left\langle H
\right\rangle _\Phi ^{\left( 2 \right)} ,  \label{susyII-50}
\end{equation}
where $\left\langle H \right\rangle _0 = \left\langle 0 \right|H\left| 0
\right\rangle$, and the $\left\langle H \right\rangle _\Phi ^{\left( 1
\right)}$ and $\left\langle H \right\rangle _\Phi ^{\left( 2 \right)}$ terms
are given by 

\begin{equation}
\left\langle H \right\rangle _\Phi ^{\left( 1 \right)}  =  - \frac{1}{{2\alpha }}\left\langle \Phi  \right|\int {d^3 x} \Pi _i \frac{{\nabla ^2 }}{{\left( {\nabla ^2  - {\raise0.5ex\hbox{$\scriptstyle \beta $}\kern-0.1em/\kern-0.15em
\lower0.25ex\hbox{$\scriptstyle \alpha $}}} \right)}}\Pi ^i \left| \Phi  \right\rangle,  \label{susyII-55}
\end{equation}

\begin{equation}
\left\langle H \right\rangle _\Phi ^{\left( 2 \right)}  = \frac{{m^2 }}{{2\alpha }}\left\langle \Phi  \right|\int {d^3 x} \Pi _i \frac{1}{{\left( {\nabla ^2  - {\raise0.5ex\hbox{$\scriptstyle \beta $}
\kern-0.1em/\kern-0.15em
\lower0.25ex\hbox{$\scriptstyle \alpha $}}} \right)}}\Pi ^i \left| \Phi  \right\rangle.  \label{susyII-60}
\end{equation}

Using Eq.(\ref{susyII-45}), we see that the potential for two opposite charges
located at $\mathbf{y}$ and $\mathbf{y^{\prime }}$ takes the form 

\begin{eqnarray}
V =  -\frac{{Q^2 }}{{4\pi a }}\frac{{e^{ - \sqrt {{\raise0.5ex\hbox{$\scriptstyle b $}\kern-0.1em/\kern-0.15em
\lower0.25ex\hbox{$\scriptstyle a $}}}\  L} }}{L}+ \frac{{Q^2 m^2 }}{{8\pi a }}\ln \left( {1 + \frac{{\Lambda ^2 }}{{{\raise0.5ex\hbox{$\scriptstyle b $}\kern-0.1em/\kern-0.15em\lower0.25ex\hbox{$\scriptstyle a $}}}}} \right)L,  \label{susyII-65}
\end{eqnarray}
where $\Lambda$ is a cutoff, $|\mathbf{y}-\mathbf{y}^{\prime }|\equiv L$,
$a = \gamma  - 32{t_{00}} + 2\frac{{{A_4}}}{{{A_2}}}$ and $b = \gamma {M^2} - 32{t_{00}}{m^2}$.
At this stage of the calculations, we must decide about the choice of the cutoff, $\Lambda$.
Following our chain of definitions for $A_1$, $A_2$, $A_3$, $A_4$, $a$, $b$, and $\gamma$, it is readily seen that the only pole that corresponds to a physical mass is exactly the photino mass, previously given in
eq. (\ref{sp20}). This means that the inter particle potential above makes sense only for distances above the Compton wavelength of the photino, ${\lambda _{photino}} \equiv m_{photino}^{ - 1}$. We then are naturally lead to make the identification $\Lambda  = {m_{photino}}$. So, our conclusion is that, whenever the pair particle-antiparticle is in static interaction at a regime of distances $r > {\lambda _{photino}}$, the form of $V$ as
given in eq. (\ref{susyII-65}) can be consistently taken. Then, with this identification, the potential of Eq. (\ref{susyII-65}) takes the form
\begin{equation}
V =  -\frac{{Q^2 }}{{4\pi a }}\frac{{e^{ - \sqrt {{\raise0.5ex\hbox{$\scriptstyle b $}\kern-0.1em/\kern-0.15em
\lower0.25ex\hbox{$\scriptstyle a $}}}\  L} }}{L}+ \frac{{Q^2 m^2 }}{{8\pi a }}\ln \left( {1 + \frac{{m_{photino}^2 }}{{{\raise0.5ex\hbox{$\scriptstyle b $}\kern-0.1em/\kern-0.15em\lower0.25ex\hbox{$\scriptstyle a $}}}}} \right)L.
\label{susyII-70}
\end{equation}

It is appropriate to observe the presence of a finite string tension in Eq. (\ref{susyII-70}).

\section{Concluding remarks}

As mentioned in the Introduction of the present contribution, there are in the literature that concerns LSV a number of approaches that contemplate the introduction of SUSY in connection with the breaking of relativistic covariance in the sense of the so-called particle transformations.

The present work a stream of investigation whose approach basically consists in assuming that LSV takes place in an environment dominated by SUSY, and we adopt the viewpoint that the bossing background usually adopted to realize the breaking of Lorentz symmetry is part of a whole set-up with fermionic SUSY partners. We then claim that LSV takes place through specific SUSY multiplets, so that the usual $k_AF$ and $k_F$-terms are accompanied by SUSY fermionic partners; in short, the background tensors that parametrize LSV are components of specific superfields.

In this paper, our main goal is to point out the salient aspects of the $k_F$-type LSV in association with an $N=1$-$D=4$-SUSY, focusing specially on the background condensates that show up along with the $(k_F)_{\mu\nu\kappa\lambda}$ breaking term. The pattern of breaking is, in the present situation, much richer than the similar inspection carried out previously in the paper of Ref. \cite{susyI}.

Particularly, the SUSY scenario for the $k_F$-LSV reveals that:
\begin{description}
\item[(i)]  The photino mass depends now not only on the bossing background (in this case, the scalar $S$) but also on the condensate $\theta  = \bar ZZ$:
\begin{equation}
m_{photino}^2 =  - \frac{{{C^{\left( 1 \right)}}}}{{16{\theta ^2}}}, \label{susyII-70}
\end{equation}
as given in eq. (\ref{sp20}). This now means that the $\theta$-condensate ($\theta$ has canonical dimension of $mass^{-1}$) may be estimated if we take the photino mass in the TeV-scale. Recalling the experimental bounds on the components of $k_{F}$ (and then on the components of the vector ${\xi ^\mu }$) \cite{Kostelecky:2008ts}, and the expression for ${C^{\left( 1 \right)}}$ in eq. (\ref{sp10}), it turns out that effectively only the condensate $\theta$ fixes the photino mass;  ${C^{\left( 1 \right)}}$ is actually of ${\cal O}(1)$. So, for a photino in the TeV-region, the condensate $\theta$ is estimated of ${\cal O}(TeV^{-1})$, corresponding then to a sort of length in the sub-millimetric scale. This result should be further exploited for it may point to an explicit SUSY breaking at an accelerator regime.

\item[(ii)] It is also remarkable to notice that, like in the $k_{AF}$-case (Carroll-Field-Jackiw), the photon dispersion relation does not receive contributions from SUSY. This feature is then common to both, $k_{AF}$- and $k_F$- cases. 

\item[(iii)] The effective photonic action is now given in terms of $\partial F$-terms, showing that, with respect to the $k_{AF}$-case, it dominates for high-energy photons and is less significant for lower frequencies.

\item[(iv)] The effects of the supersymmetric background fermion condensates are, moreover, felt through of the photonic action. It is therefore not surprising that they become manifest in the interaction energy for the effective theory. In fact, we have obtained the effective theory for the condensed phase and computed the interaction energy between two static charges, in order to test the confinement versus screening properties of our effective model. Interestingly, we explicitly shown that the static potential profile contains an Yukawa term and a linear term, leading to the confinement of static charges.

\end{description}

Finally, we would like to comment that we could also inspect this very same model (the $k_F$-model) by considering the ${\xi ^\mu }$-vector not given by the scalar supermultiplet as the $4$-gradient of $S$. We could rather suppose that ${\xi ^\mu }$ is placed in a (non-gauge) vector multiplet of $N=1-D=4-$ SUSY, which would introduce a richer fermionic background. Moreover, $\xi^\mu$ would in this case become a complete vector, with a transverse part in addition to its gradient (longitudinal component). A wider class of condensates would emerge in such a situation and this might have an interesting consequence specially in the photon dispersion relations, always very sensitive to the particular choice of the multiplet that accommodates the background yielding LSV. We are already concentrating efforts in this direction and we shall be reporting our results in a forthcoming paper to better understand the influence of the particular supersymmetric structure on the physics of LSV.

\section{acknowledgements}

The authors (HB, LDB and JAH-N) express their gratitude to the Conselho Nacional de Desenvolvimento Cient\'{\i}fico e Tecnol\'{o}gico (CNPq-Brazil) for the financial support. P. G. was partially supported by FONDECYT (Chile) grant 1130426, DGIP (UTFSM) internal project USM 111458.

\section{Appendix A}

Below, we collect the 3 pieces of our component-field action corresponding to Eq. (\ref{cbpf05})   in terms of ($2$-component) Weyl spinors:
\begin{eqnarray}
S_{boson}= \int d^4x \Big[ &&D^2 (32|G|^2+16\partial_\mu S \partial_\mu S^*)+8iD F^{\mu\nu}(\partial_\mu S\partial_\nu S^*- \partial_\mu S^*\partial_\nu S) \nonumber\\
&&-8F^{\mu\kappa}F_{\kappa}^{\,\,\,\,\nu}(\partial_\mu S\partial_\nu S^*+ \partial_\mu S^*\partial_\nu S)-4F^{\mu\nu}F_{\mu\nu} \partial_\alpha S \partial_\alpha S^* \,\,\,\Big], \nonumber\\
S_{{ferm}}= \int d^4 x \Big{[}&& \,\frac{1}{2}\partial_\lambda \zeta \sigma^\mu \partial_\mu\bar{\zeta}\lambda \sigma^\lambda\bar{\lambda}+\frac{1}{2}\partial_\lambda \zeta\sigma^\mu \bar{\lambda}\lambda \sigma^\lambda\partial_\mu \bar{\zeta}+2\partial_\mu \zeta \partial^\mu \lambda \bar{\zeta}\bar{\lambda}+\notag\\
&&-\frac{1}{2}\partial_\lambda \zeta \sigma^\lambda \partial_\mu \bar{\zeta} \lambda \sigma^\mu \bar{\lambda}-2\partial_\lambda \zeta \sigma^\lambda \bar{\sigma}^\mu \partial_\mu \lambda \bar{\zeta}\bar{\lambda}-\frac{1}{2}\lambda \sigma^\lambda \bar{\sigma}^\mu \partial_\lambda \zeta \bar{\zeta} \partial_\mu \bar{\lambda}+\notag\\
&&-\frac{1}{2}\zeta \lambda \bar{\zeta}\Box \bar{\lambda}-\zeta \lambda \partial_\mu \bar{\zeta}\bar{\sigma}^\mu \sigma^\tau \partial_\tau \bar{\lambda}+\frac{1}{2} \zeta \lambda \partial_\mu \bar{\lambda}\bar{\sigma}^\nu \sigma^\mu \partial_\nu \bar{\zeta}+\notag\\
&&+\frac{1}{2}\partial_\mu \zeta \sigma^\mu \bar{\sigma}^\nu \lambda \bar{\zeta}\partial_\nu \bar{\lambda}-\frac{1}{2\sqrt{2}}\zeta \Box \lambda \bar{\zeta}\bar{\lambda}-\frac{1}{2}\zeta\partial_\nu \lambda\partial_\mu \bar{\lambda}\bar{\sigma}^\mu \sigma^\nu \bar{\zeta}+\notag\\
&&-\frac{1}{2}\zeta \partial_\nu \lambda \partial_\mu \bar{\zeta}\bar{\sigma}^\nu \sigma^\mu \bar{\lambda}+\zeta \partial_\mu \lambda \bar{\zeta} \partial^\mu \bar{\lambda}-2\zeta \sigma^\mu \partial_\mu \bar{\lambda}\bar{\zeta}\bar{\sigma}^\nu \partial_\nu \lambda+h.c. \Big],\notag\\
S_{mixing}= \int d^4x \big[ &&-4iD^2 \zeta\sigma^\mu \partial_\mu \bar{\zeta}-2\sqrt{2}i D G^* \zeta \sigma^\mu\partial_\mu \bar{\lambda}+2\sqrt{2}D \partial_\nu \lambda \sigma^\nu \bar{\sigma}^\mu \zeta \partial_\mu S^*+ \notag\\
&&2D\zeta \sigma^\nu \partial_\mu \bar{\zeta} F_\nu^{\,\,\,\mu}+i D \epsilon^{\tau\rho\mu\alpha} \zeta \sigma_\alpha \partial_\mu \bar{\zeta} F_{\tau \rho}+\sqrt{2}G^* \zeta \sigma^\mu  \partial_\nu \bar{\lambda} F_\mu^{\,\,\,\nu}+\notag\\
&&\frac{i}{\sqrt{2}} G^* \epsilon^{\tau\rho\mu\alpha} \zeta \sigma_\alpha \partial_\mu \bar{\lambda} F_{\tau \rho}+\sqrt{2}i \zeta \sigma^\tau \bar{\sigma}^\nu \partial_\nu \lambda \partial_\mu S^* F_\tau^{\,\,\,\mu}+\nonumber\\
&&-\frac{1}{\sqrt{2}} \epsilon^{\tau\rho\mu\alpha} \zeta \sigma_\alpha \bar{\sigma}^\nu \partial_\mu S^* \partial_\nu \lambda F_{\tau\rho}-4\sqrt{2}i G^*D \zeta \sigma^\mu \partial_\mu \bar{\lambda}+\nonumber\\
&&+2\sqrt{2}D \zeta \partial_\mu \lambda \partial^\mu S^*-\frac{i}{\sqrt{2}}\epsilon^{\mu\nu\kappa\tau} \zeta \partial_\tau \lambda \partial_\mu S^* F_{\nu\kappa}+\frac{1}{2\sqrt{2}}\epsilon^{\mu\nu\kappa\tau} \zeta \partial_\tau \lambda  \partial_\mu S^* F_{\nu\kappa}+\nonumber\\
&&-4iD^2 \bar{\zeta}\bar{\sigma}^\mu \partial_\mu \zeta-2D\bar{\zeta}\bar{\sigma}^\nu \partial_\mu \zeta F_\nu^{\,\,\,\mu}+iD \epsilon^{\nu\kappa\mu\alpha}\bar{\zeta}\bar{\sigma}_\alpha \partial_\mu \zeta F_{\nu\kappa}+\nonumber\\
&&+2\sqrt{2}i D G^* \partial_\mu \zeta \sigma^\mu \bar{\lambda}+2D\partial_\mu \sigma^\tau \bar{\zeta} F_\tau^{\,\,\,\mu}+iD \epsilon^{\tau\rho\mu\alpha} \partial_\mu \zeta \sigma_\alpha \bar{\zeta} F_{\tau\rho}+\nonumber\\
&&2\partial_\mu \zeta \sigma^\tau \bar{\sigma}^{\nu\kappa} \bar{\zeta} F_{\nu\kappa} 
F_\tau^{\,\,\,\mu}+i \epsilon^{\tau\rho\mu\alpha} \partial_\mu \zeta \sigma_\alpha \bar{\sigma}^{\nu\kappa}\bar{\zeta}F_{\tau\rho} F_{\tau\rho}+\sqrt{2}G^*\partial_\mu \zeta\sigma^\tau \bar{\lambda}F_{\tau}^{\,\,\,\mu}\nonumber\\
&&-2iG\partial_\nu \lambda \sigma^\nu \bar{\sigma}^\mu \lambda \partial_\mu S^*+4\sqrt{2}iGD \partial_\mu \lambda \sigma^\mu \bar{\zeta}-2\sqrt{2}iGD \bar{\zeta}\bar{\sigma}^\mu \partial_\mu \lambda+\nonumber\\
&&-\sqrt{2}G \bar{\zeta}\bar{\sigma}^\mu \partial_\tau \lambda F_\mu^{\,\,\,\tau}+\dfrac{i}{\sqrt{2}}G \epsilon^{\mu\nu\tau\alpha}\bar{\zeta}\bar{\sigma}_\alpha \partial_\tau \lambda F_{\mu\nu}-2i|G|^2 \bar{\lambda}\bar{\sigma}^\mu \partial_\mu \lambda+\nonumber\\
&&+\frac{i}{\sqrt{2}} G^* \epsilon^{\tau\rho\mu\alpha} \partial_\mu \zeta \sigma_\alpha \bar{\lambda} F_{\tau\rho}-2\sqrt{2}i G D \lambda \sigma^\mu \partial_\mu \bar{\zeta}+2i|G|^2 \lambda \sigma^\mu \partial_\mu \bar{\lambda}+\nonumber\\
&&2\sqrt{2}D \partial_\mu S \bar{\zeta} \partial^\mu \bar{\lambda}+\sqrt{2}i \partial_\mu (\bar{\lambda}\bar{\zeta}) \partial_\lambda S F^{\lambda\mu} -\frac{1}{\sqrt{2}}\epsilon^{\mu\lambda\tau\rho}\partial_\mu (\bar{\lambda}\bar{\zeta})\partial_\lambda S F_{\tau\rho}+\nonumber\\
&&-2\sqrt{2}D \bar{\lambda}\partial_\mu \bar{\zeta}\partial^\mu S+2\sqrt{2}D \bar{\zeta}\bar{\sigma}^\mu \sigma^\nu \partial_\nu \bar{\lambda} \partial_\mu S-\sqrt{2}i\bar{\zeta}\bar{\sigma}^\nu \sigma^\mu \partial_\mu \bar{\lambda} F_{\nu\lambda} \partial^\lambda S+\nonumber\\
&&-\frac{1}{\sqrt{2}}\epsilon^{\nu\kappa\lambda\alpha} \bar{\zeta}\bar{\sigma}_\alpha \sigma^\mu \partial_\mu \bar{\lambda} F_{\nu\kappa}\partial_\lambda S+2 G^* \bar{\lambda} \bar{\sigma}^\mu \sigma^\nu \partial_\nu \bar{\lambda} \partial_\mu S-2\sqrt{2} \partial_\mu S \partial^\mu D \bar{\zeta} \bar{\lambda}+\nonumber\\
&&+\sqrt{2}D \partial_\nu \zeta \sigma^\nu \bar{\mu} \lambda\partial_\mu S^*-\frac{i}{\sqrt{2}}\bar{\sigma}^\lambda \partial_\lambda \zeta \lambda \sigma^\nu \partial_\mu S^* F_\nu^{\,\,\,\mu}+\frac{1}{2\sqrt{2}}\epsilon^{\mu\nu\kappa\alpha}\lambda \sigma_\alpha \bar{\sigma}^\lambda \partial_\lambda \zeta \partial_\lambda \zeta \partial_\mu S^* F_{\nu\kappa}+\nonumber\\
&&-\frac{1}{2}\partial_\lambda \zeta \sigma^\lambda \bar{\lambda} \lambda \sigma^\mu \partial_\mu \bar{\zeta}+\sqrt{2}D \lambda \sigma^\lambda \bar{\sigma}^\mu \partial_\lambda \zeta \partial_\mu S^*-\frac{i}{\sqrt{2}} \partial_\lambda \zeta \sigma^\nu \bar{\lambda} \lambda \partial_\mu S^* F_\nu^{\,\,\,\mu}+\nonumber\\
&&-\frac{1}{2\sqrt{2}}\epsilon^{\mu\nu\kappa\alpha} \partial_\lambda \zeta \sigma_\alpha \bar{\sigma}^\lambda \lambda \partial_\mu S^* F_{\nu\kappa}+h.c \big]. 
\end{eqnarray}

\section{Appendix B}

To render more fluent the text of Section III, we present, in this Appendix, the full expression and details related to the ${\beta}^2$-term that yields our final expression for the action (\ref{cbpf05}) after the $D$-auxiliary field is eliminated in favor of its equation of motion.
\begin{eqnarray}
\beta^2&=&\bar{\Lambda} \Big{(}\tilde{a}+\tilde{b}\gamma_5+\tilde{\omega}^\rho\gamma_\rho \gamma_5\Big{)}\Lambda + \nonumber\\
&16& m_{\mu\nu}F^{\mu\nu} \Big{[}10\sqrt{2} Re(\partial_\mu S)(\bar{Z}\partial_\mu \Lambda)-8\sqrt{2}i Re(\partial_\mu S) (\bar{Z}\Sigma^{\mu\nu}\partial_\nu \Lambda)
8\sqrt{2}  Im(\partial_\mu S)(\bar{Z}\Sigma^{\mu\nu}\gamma_5 \partial_\nu \Lambda)+10\sqrt{2}i Im(\partial_\mu S)(\bar{Z}\gamma_5 \partial^\mu \Lambda)\Big{]} \nonumber\\
&+&256 m_{\mu\nu}m_{\alpha\beta}F^{\mu\nu}F^{\alpha\beta},
\end{eqnarray}
where the operators $\tilde{a}$, $\tilde{b}$ and $\tilde{w}^\rho$ are defined as: 
\begin{equation}
\tilde{a}=42  \theta s^{\alpha\beta}\Box \omega_{\alpha\beta}+84i \tau Re(\partial^\alpha S)Im(\partial^\beta S)\Box \omega_{\alpha\beta} +8 \theta s\Box +16i\tau Re(\partial_\rho S)Im(\partial^\rho S) \Box, 
\end{equation}
\begin{equation}
\tilde{b}=42  \tau s^{\alpha\beta}\Box \omega_{\alpha\beta}+84i \theta Re(\partial^\alpha S)Im(\partial^\beta S)\Box \omega_{\alpha\beta} + 8 \tau s\Box +16i\theta Re(\partial_\rho S)Im(\partial^\rho S),
\end{equation}
\begin{eqnarray}
\tilde{w}^{\rho}&=&-50 C^\rho t^{\alpha\beta}\Box \omega_{\alpha\beta}+40 t^{\alpha}_{\,\,\theta}(C^\beta \eta^{\theta\rho}-C^\theta \eta^{\beta\rho})\Box \omega_{\alpha\beta} \nonumber\\
&+&40[Im(\partial^\alpha S)Re(\partial^\nu S)-Re(\partial^\alpha S)Im(\partial^\nu S)]C^\theta \epsilon_{\theta\nu}^{\,\,\,\,\,\,\,\beta\rho}\Box \omega_{\alpha\beta}+8 C_{\theta}(r^{\theta\rho\alpha\beta}+s^{\theta\alpha\beta\rho})\Box \omega_{\alpha\beta} \label{app05}.
\end{eqnarray}
Expression (\ref{app05}) may be rewritten as
\begin{equation}
\tilde{w}^\rho=-4(1+8 \partial_\kappa S \partial^\kappa S^*)]C_\alpha d^{\alpha\rho}.
\end{equation}
To get the last line we have used:
\begin{equation}
r^{\theta\alpha\beta\rho}=(\eta^{\theta\alpha}\epsilon^{\nu\mu\beta\rho}+\eta^{\rho\alpha}\epsilon^{\nu\mu\beta\theta}+\eta^{\rho\beta}\epsilon^{\nu\mu\alpha\theta}+\eta^{\theta\beta}\epsilon^{\nu\mu\alpha\rho})Re(\partial_\nu S) Im(\partial_\mu S),
\end{equation}
\begin{equation}
u^{\theta\rho\alpha\beta}=2t^{\theta\rho}\eta^{\alpha\beta}-2t^{\theta\alpha}\eta^{\beta\rho}-2t^{\beta\rho}\eta^{\theta\alpha}+t^{\alpha\beta}\eta^{\theta\rho}+t(2\eta^{\theta\alpha}\eta^{\beta\rho}-\eta^{\alpha\beta}\eta^{\theta\rho}),
\end{equation}
\begin{equation}
s_{\alpha\beta}=Im(\partial_\alpha S)Im(\partial_\beta S)-Re(\partial_\alpha S)Re(\partial_\beta S),
\end{equation}
\begin{equation}
t_{\alpha\beta}=Im(\partial_\alpha S)Im(\partial_\beta S)+Re(\partial_\alpha S)Re(\partial_\beta S),
\end{equation}
\begin{equation}
t=\eta^{\alpha\beta}t_{\alpha\beta},
\end{equation}
\begin{equation}
s=\eta^{\alpha\beta}s_{\alpha\beta},
\end{equation}
and
\begin{equation}
w_{\alpha\beta}=\frac{\partial_\alpha \partial_\beta}{\Box}.
\end{equation}

\end{document}